\documentclass[aop,preprint]{imsart}

\RequirePackage[OT1]{fontenc}
\RequirePackage{amsthm,amsmath,amsfonts}
  \usepackage{epsfig} 
\usepackage{graphicx}
\usepackage{lineno,hyperref}
\usepackage{float, comment}
\usepackage{epstopdf}

\RequirePackage[numbers]{natbib}
\RequirePackage[colorlinks,citecolor=blue,urlcolor=blue]{hyperref}

\usepackage{amsmath,amsxtra,latexsym,amsthm, amssymb, amscd}

\startlocaldefs
\numberwithin{equation}{section}
\theoremstyle{plain}

\theoremstyle{definition}

\endlocaldefs

\allowdisplaybreaks

\begin{document}

\begin{frontmatter}

\title{Mathematical Models for Fish Schooling}
\runtitle{Fish Schooling}

\begin{aug}

\author{Linh Thi Hoai Nguyen, T\^{o}n Vi$\hat{\d{e}}$t  T\d{a},
Atsushi Yagi}

\ead[label=e1]{tavietton[at]agr.kyushu-u.ac.jp}
\ead[label=e2]{nguyen.thi.hoai.linh.578[at]bioreg.kyushu-u.ac.jp}
\ead[label=e3]{atsushi-yagi[at]ist.osaka-u.ac.jp}

\address{
Linh Thi Hoai Nguyen\\
Department of Immunobiology and Neuroscience\\
Medical Institute of Bioregulation, Kyushu University\\
3-1-1 Maidashi, Higashi-ku, Fukuoka 812-8582, JAPAN\\
\printead{e2}\\
\\
T\^{o}n Vi$\hat{\d{e}}$t  T\d{a}\\
Center for Promotion of International Education and Research 
\\
Faculty of Agriculture, 
Kyushu University
\\
6-10-1 Hakozaki, Higashi-ku, Fukuoka 812-8581, JAPAN\\
\printead{e1}\\
\\
Atsushi Yagi\\
Department of Applied Physics\\
 Graduate School of Engineering, Osaka University\\
1-5 Yamadaoka, Suita, Osaka 565-0871, JAPAN\\
\printead{e3}}

\end{aug}

\begin{abstract}
This note reviews our mathematical models for fish schooling, considered in free space, and in space with obstacle and food resource. These models are performed by stochastic differential equations or stochastic partial differential equations. We then present an example for the model in the last case.
\end{abstract}


\end{frontmatter}

\section{Introduction}  \label{introduction}

Swarming behavior of a fish school consisting of a large number of individuals often surprises us. They swim coherently matching their velocity without collision and maintaining a constant scale of school, even though they have only moderate ability of information processing and of execution of programming. 

Several  mathematical models are presented on the basis of experimental results concerning interactions between nearby mates which are rather simple. Vicsek-Czir\'{o}k-Ben Jacob-Cohen-Shochet (\cite{Vicsek1995}) introduced a simple difference model, assuming that each particle is driven with a constant absolute velocity, and  chooses its new heading to be the average of those of nearby particles located within a unit distance. 
Oboshi-Kato-Mutoh-Itoh (\cite{Oboshi2002}) presented another difference model in which an individual selects one basic behavioral pattern from four  based on the distance between it and its nearest neighbor.  Olfati-Saber (\cite{RezaMultiAgent}) and D\'{}Orsogna-Chuang-Bertozzi-Chayes (\cite{DOrsogna2006}) constructed a deterministic differential model using a generalized Morse and attractive/repulsive potential functions, respectively.
For more references, we refer the reader to \cite{Aoki1982}, \cite{Cucker1}--\cite{CS1},\cite{Huth1992}--\cite{LinhTonYagi},\cite{Reynolds1987}--\cite{Uchitane2012}, and \cite{Zienkiewicz}.

In this paper, we first review two of our mathematical models for fish schooling.  The first one describes fish schooling in free space,  performed by stochastic differential equations (SDEs) (\cite{Uchitane2012}).  Meanwhile, the second one describes foraging behavior of fish schools in noisy environment with  obstacle and food resource. It is governed by stochastic partial differential equations (\cite{Nguyen2017}). We then give an example for the last model.

The organization of  the paper is as follows. In Sections  \ref{sec2} and \ref{sec3}, we review the two models.  
 Section \ref{sec4} presents an example.

\section{Fish schooling model based on five components in free spaces}\label{sec2}
%------------------------------- Subsection SDE model in free space 

In \cite{Uchitane2012}, we introduced five components constructing a SDE model for fish schooling in free space. These components are based on the following three local rules of Camazine-Deneubourg-Franks-Sneyd-Theraulaz-Bonabeau 
 (\cite{Camazine2001}):
\begin{enumerate}
\item [(R1)] The school has no leaders and each fish follows the same behavioral rules.
\item [(R2)] To decide where to move, each fish uses some form of weighted average of the position and orientation of its nearest neighbors.
\item [(R3)] There is a degree of uncertainty in the individual's behavior that reflects both the imperfect information-gathering ability of a fish and the imperfect execution of the fish's actions.
\end{enumerate}

Let us recall the model. Consider a group of $N$ fish.  
Each is regarded as a  particle moving in the free space $\mathbb R^d$ $(d=1,2,3\dots)$ with norm $\|\cdot\|$. Denote by ${x}_i(t)$ and ${v}_i(t)$ $(i=1,2\dots N)$ the position and the velocity, respectively, of the $i$-th individual at time $t$. 

The particle-particle and particle-environment interactions consist of
\begin{enumerate}
  \item Attraction force. When two particles $i$ and $j$ are far from each other, both would move toward each other. In our model, this force is a generalization of the inverse-square law of universal gravitation:
     $$ \frac{-\alpha r^p({x}_i-{x}_j)}{\|{x}_i-{x}_j\|^p},$$ 
where $1<p<\infty$ and $r>0$ are constants, $\alpha$ a coefficient of attraction among individuals.
 \item Repulsion force. When two particles $i$ and $j$ are close enough, both would move far from each other. This force is a generalization of the Van der Waals forces:
     $$ \frac{\alpha r^q({x}_i-{x}_j)}{\|{x}_i-{x}_j\|^q},$$ 
where $1<p<q<\infty$ is a fixed exponent.
\item Alignment or velocity matching. 
The velocity matching of the particle $i$ to the particle $j$ also has a similar weight depending on the distance $\|{x}_i-{x}_j\|$:
$$-\beta\left(\frac{r^p}{\|{x}_i-{x}_j\|^p}+\frac{r^q}{\|{x}_i-{x}_j\|^q} \right)({v}_i-{v}_j),$$
where $\beta$ is a coefficient of velocity matching among individuals.
\item Reaction to the environment. The  individual $i$ react to the environment a force $F_i({x}_i,{v}_i).$ 
\item Noise. All particles are subject to random factors or noise.
\end{enumerate}
These five components form our SDE model of the form
\begin{equation}\label{eq0}
\begin{aligned}
\begin{cases}
d{x}_i(t)=&{v}_idt+\sigma_idw_i(t), \hspace{1cm} i=1,2,\dots N,\\
d{v}_i(t)=&\Big\{-\alpha\sum\limits_{j=1,j\ne i}^N\left(\frac{r^p}{\|{x}_i-{x}_j\|^p}-\frac{r^q}{\|{x}_i-{x}_j\|^q} \right)({x}_i-{x}_j)\\
  &\quad -\beta\sum\limits_{j=1,j\ne i}^N\left(\frac{r^p}{\|{x}_i-{x}_j\|^p}+\frac{r^q}{\|{x}_i-{x}_j\|^q} \right)({v}_i-{v}_j)\\
&\quad +F_i({x}_i,{v}_i)\Big\}dt, \hspace{1cm} i=1,2,\dots N.
\end{cases}
\end{aligned}
\end{equation}

 The first equation of \eqref{eq0} is a stochastic equation for the unknown ${x}_i(t)$, where $\sigma_idw_i$  denotes a stochastic differentiation of a $d$-dimensional independent Brownian motion defined in a filtered probability space.  The second one is a deterministic equation for the unknown ${v}_i(t)$.
If the fish $i$ is far from the fish $j$, i.e., $\|{x}_i-{x}_j\|>r,$ then it would move toward the other due to the attraction force. To the contrary, if they are close enough, i.e., $\|{x}_i-{x}_j\|<r,$ then both would avoid collision with each other due to the repulsive force. The quantity $r$ therefore plays as the critical distance. 

The reader can find some mathematical results for the system \eqref{eq0} in \cite{Nguyen2014,Uchitane2012}.

An advantage of using SDE models like \eqref{eq0} may be the easiness of mathematical treatments. One can utilize the well-developed theory of SDEs and the numerical methods. Its flexibility may be another advantage. As seen in the next section, we make a new model by introducing suitable functions $F_i$ in \eqref{eq0}.

\section{Fish schooling model in spaces with obstacle and food resource}   \label{sec3}

In  \cite{LinhTonYagi}, we introduced a fish schooling model in spaces with obstacle. For this model, in addition to the three local rules (a)--(c) in the previous section, we newly presented a local rule of obstacle avoidance for individual fish:
\begin{enumerate}
\item[(R4)] Each fish executes an action for avoiding obstacle according to the reflection law of velocity with a weight depending on distance.
\end{enumerate}

Furthermore, we investigated foraging behavior of fish schools in noisy environment with  obstacle and food resource in \cite{Nguyen2017}. 
 Consider a fish school moving in a free or limited space to forage for food. The position of food resource is fixed in the space. Fish and food may be separated by obstacles in the sense that the school cannot move to the food in a straightforward way. 

To construct a mathematical model for foraging behavior, we introduced for the first time a local rule for foraging:
\begin{enumerate}
\item[(R5)] Each fish is sensitive to the gradient of potential formed by scent which is emitted by food, and has tendency to move into a higher direction.
\end{enumerate}

Let us first recall the mathematical formulation of  the local rule  {\rm (R5)} by using a method of potential functions. 

Let $F$ be the density function of food resource defined in a domain $ \Omega \subset \mathbb R^d$.  Consider an elliptic equation
in $\Omega$ under the homogeneous Neumann boundary condition on $\partial \Omega$: 
\begin{equation}\label{eq2}
\begin{aligned}
\begin{cases}
-\alpha_1 \Delta X+\alpha_2 X=F(x), &  \hspace{2cm}  x\in \Omega, \\
\dfrac{\partial X}{\partial \text{\bf n}}=0, & \hspace{2cm} x \in \partial \Omega.
\end{cases}
\end{aligned}
\end{equation}   
 Here, $X(x)$ denotes the density of scent emitted by food at $x\in \Omega$. The operator $\Delta$ is the Laplace operator in $\Omega$; $\alpha_1 >0$ is a diffusion constant;  $\alpha_2 >0$ is a declining rate of $X(x)$; and  $\text{\bf n}$ denotes the exterior normal to the boundary $\partial \Omega$.  The Neumann boundary condition ensures that the domain is perfectly insulated, i.e., scent of food cannot pass through the boundary of the domain.

We regard $X$  as a potential function. Assume that the fish $i$ is at  position $x_i \in \Omega$ at some moment.  Let $G_i$  be the gradient of the potential function, i.e., 
\begin{equation} \label{eq3}
G_i(x_i)=\alpha_3\nabla X(x_i), \hspace{1cm} i=1,2\dots N,
\end{equation}
where $\alpha_3>0$ is a sensitivity constant. The fish $i$ is sensitive to $G_i$.

We are now ready to restate model equations for foraging behavior of fish schools in noisy environment with  obstacle and food resource:
\begin{equation} \label{eq4}
\begin{cases}
\begin{aligned}
  dx_i(t) = & v_i dt+\sigma_idw_i(t),   \hspace{1cm} i=1,2\dots N, \\
  dv_i(t) = &\Bigl[  - \alpha \sum\limits_{j=1,\, j\ne i}^N
    \left(\dfrac{r^p}{\|x_i - x_j\|^p}- \dfrac{r^q}{\|x_i-x_j\|^q} \right)(x_i-x_j)   \\
&\quad - \beta \sum\limits_{j=1,\, j\not=i}^N
    \left(\dfrac{r^p}{\|x_i-x_j\|^p} +\dfrac{r^q}{\|x_i-x_j\|^q} \right)(v_i-v_j)\\
&  \quad -\gamma\left(\dfrac{R^P}{\|x_i-x_i^*\|^P}+\dfrac{R^Q}{\|x_i-x_i^*\|^Q}\right) (v_i-v_i^*)\\
 & \quad +\alpha_3\nabla X(x_i)\Bigl]dt,  \hspace{1cm} i=1,2\dots N,
\end{aligned}\\
\begin{aligned}
&-\alpha_1\Delta X+\alpha_2 X= F(x),    &  \hspace{2cm}x\in \Omega, \\
&\dfrac{\partial X}{\partial \text{\bf n}}= 0,  &   \hspace{2cm}x \in \partial \Omega.
\end{aligned}
 \end{cases}
\end{equation}
Here, $\Omega$ is the domain of $\mathbb R^d$ which fish moves in; $x_i^*\in\partial \Omega$ is a point which the fish is feared to collide at; $R>0$ is a fixed distance; $\gamma>0$ is a constant; and  $1<P<Q<\infty$ are exponents. The vector $v_i^*$ is the reflection vector of vector $v_i$  with respect to obstacle $\partial\Omega$. (The fish at $x_i$ avoids $\partial\Omega$ by mathching its velocity to $v_i^*$.)

\section{An example}     \label{sec4}

In this section, we give an example  for the model \eqref{eq4} in two-dimensional space. The reader can find more results on \eqref{eq4} in \cite{Nguyen2017}.

Put a food resource at a small circle of radius $0.04$ and center  $(6, 0.1).$ 
 More precisely, the function $F$ of food resource in \eqref{eq2} has the form:
\begin{equation}    \label{foodresource}
\begin{aligned}
F(x)=\begin{cases}
50   &  \hspace{1cm} \text{if  }  x\in \{y \in \mathbb R^2: \|y- (6, 0.1)\|\leqslant 0.04\}, \\
0 &  \hspace{1cm} \text{else}.
\end{cases}
\end{aligned}
\end{equation} 
Set $\alpha_1=0.1, \alpha_2=0.2, $ and the  domain 
$$\Omega=[0, 7]\times [0, 4]\setminus \{[2, 2.5]\times[2.5,4] \cup [4.5, 5]\times [0,1.5]\},$$
 where 
$[2, 2.5]\times[2.5,4]$ and $[4.5, 5]\times [0,1.5]$ are two obstacles. By the Neumann condition in the elliptic equation  \eqref{eq2}, the scent of food  cannot pass through the boundary $\partial \Omega$.

We  set initial values and parameters for the system  \eqref{eq4} as follows. All initial positions of $25$-fish $(N=25)$ are taken randomly in the rectangle domain $[0,2]\times [3.5,4]$, meanwhile all initial velocities are null. Furthermore, $\alpha=1, \beta=0.5, \gamma=1$, $p=P=3$, $q=Q=4$, $r=0.1$, $R=0.2,$ $\alpha_3=5.5,$ and $\sigma=0.001.$

We introduce a parameter $\|v\|_{\max}=0.8$ to restrict speed of fish. If the magnitude of $v_i$ exceeds $\|v\|_{\max}$, our program would reset $v_i$ to a vector of magnitude $\|v\|_{\max}$ and same direction. That is  
\begin{equation*}
v_i(t)=
\begin{cases}
v_i(t)\qquad &\text{\rm if} \quad \|v_i(t)\|\leqslant \|v\|_{\max},\\
\frac{v_i(t)}{\|v_i(t)\|}\|v\|_{\max}\qquad &\text{\rm otherwise}.
\end{cases}
\end{equation*}

\begin{figure}[H]
\begin{center}
\includegraphics[width=6cm, height=4cm]{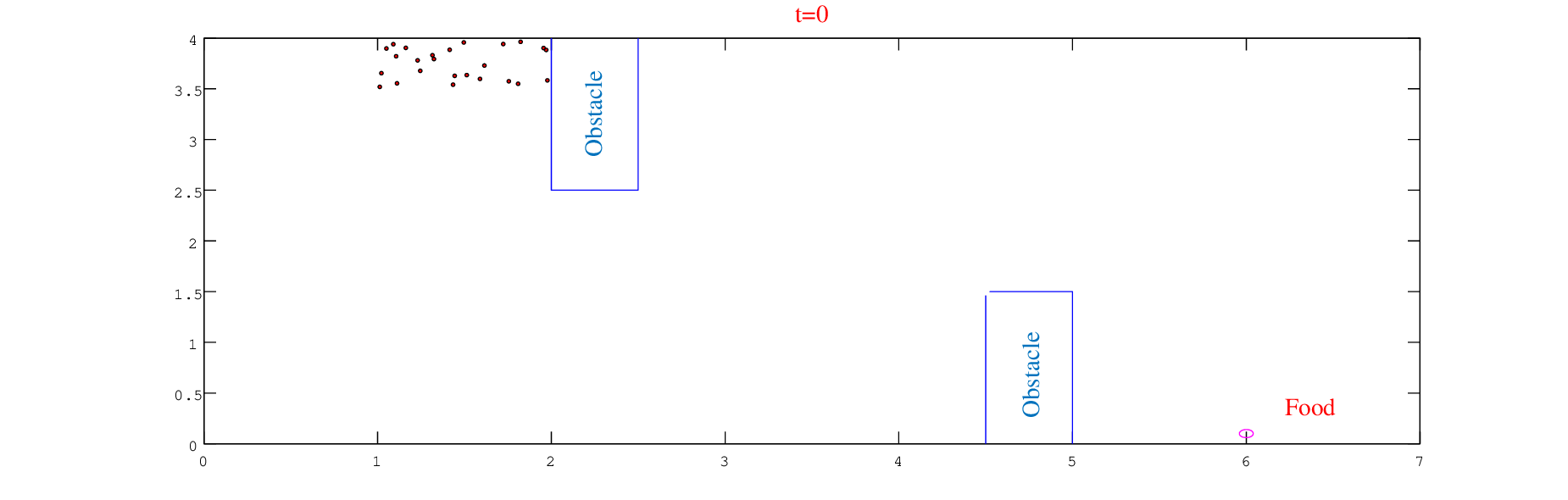} 
\includegraphics[width=6cm, height=4cm]{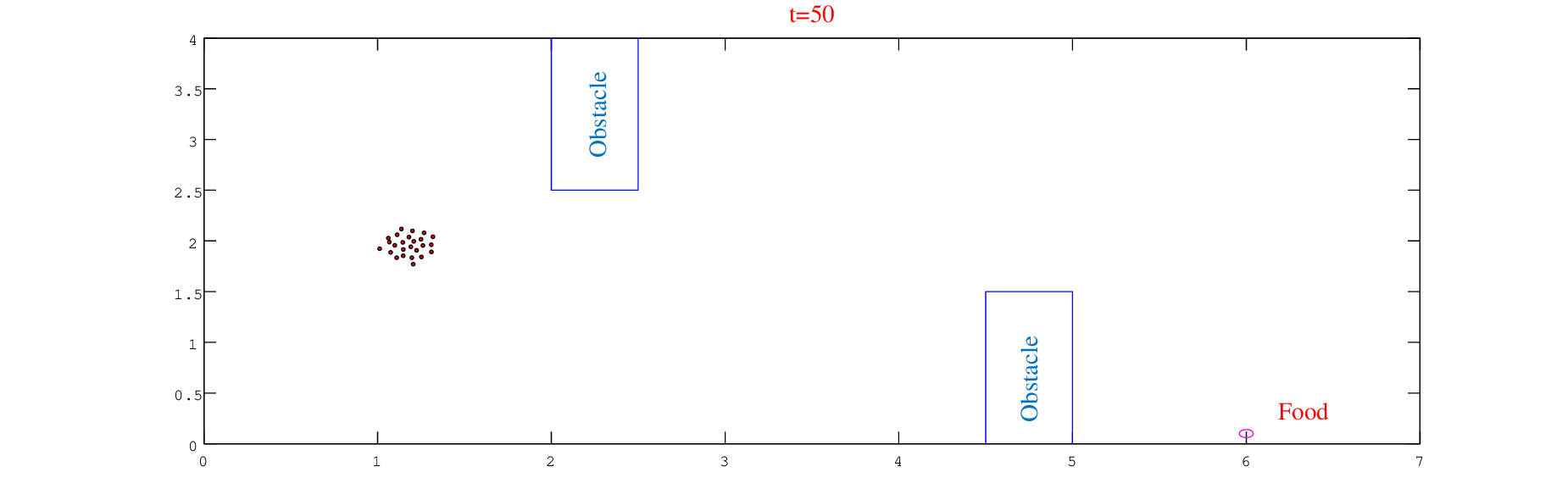} 
\includegraphics[width=6cm, height=4cm]{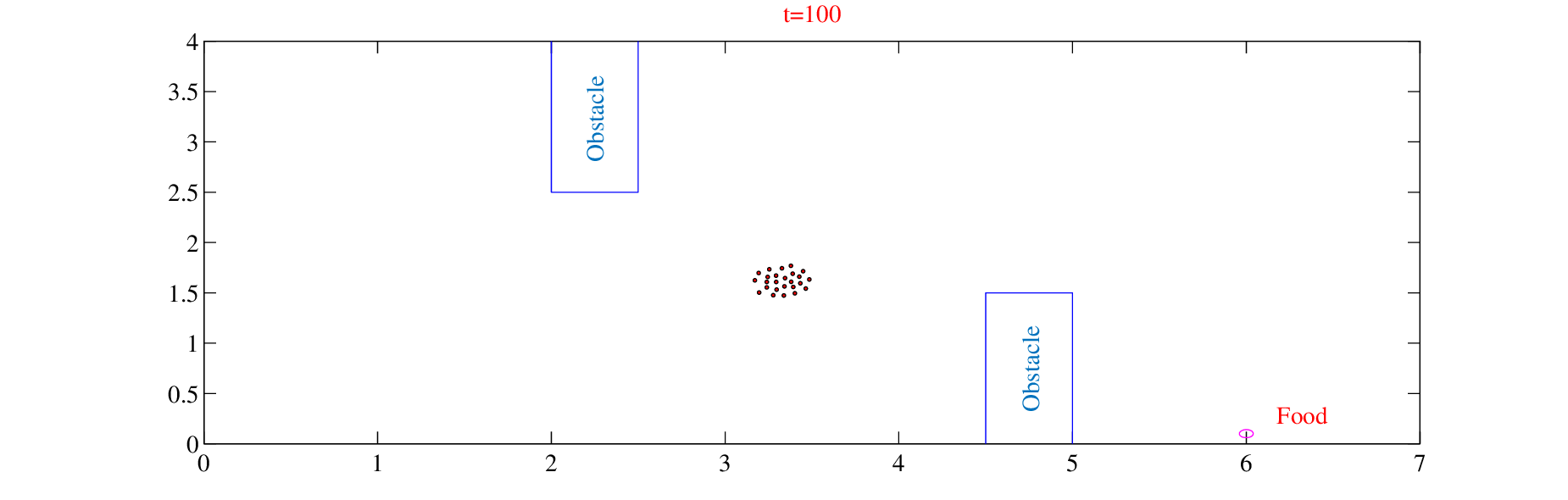} 
\includegraphics[width=6cm, height=4cm]{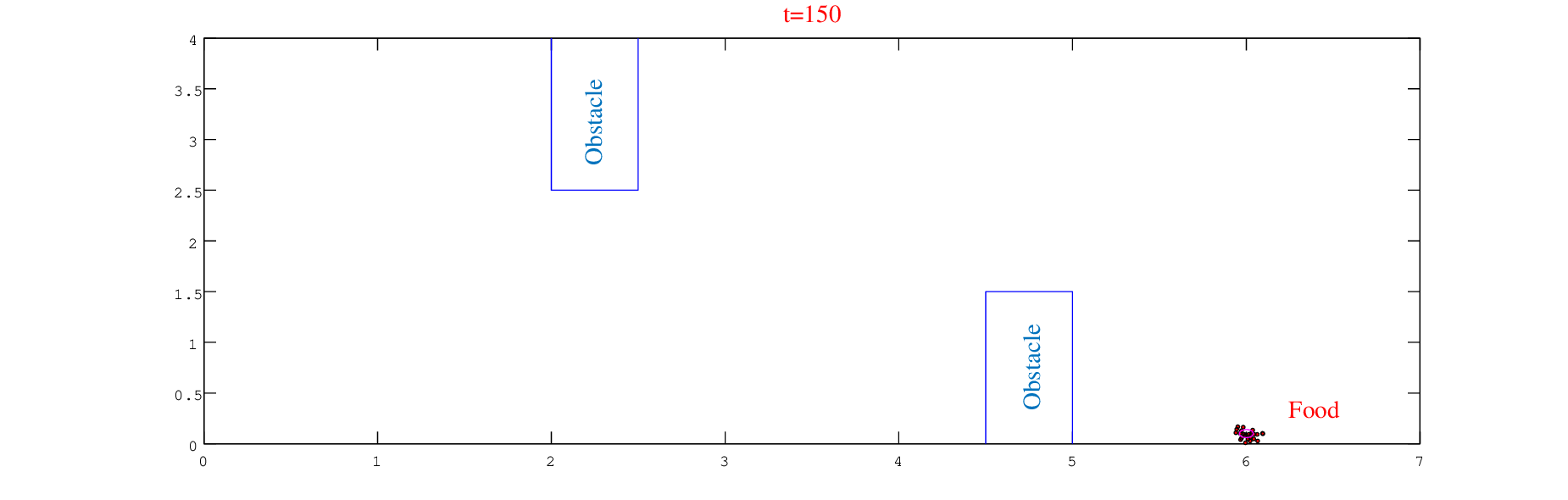} 
 \caption{A pattern of collective foraging. Positions of all 25 fish in school are plotted at four  instants. The school reaches to the food resource at $(6, 0.1)$ while maintaining its school structure.} 
  \label{Config2_position}
 \end{center}
\end{figure}

Figure  \ref{Config2_position} shows  a pattern of collective foraging. Positions of all 25 fish in school are plotted at four instants $t=0, 50, 100, 150$. The fish school reaches to the food source.

%\section*{Acknowledgments} 

\end{document}